\documentstyle[prd,aps,twocolumn]{revtex}
\begin{document}
\draft
\newcommand{\beq}{\begin{equation}}
\newcommand{\eeq}{\end{equation}}
\newcommand{\bdis}{\begin{displaymath}}
\newcommand{\edis}{\end{displaymath}}
\newcommand{\bea}{\begin{eqnarray}}
\newcommand{\eea}{\end{eqnarray}}
\newcommand{\barr}{\begin{array}}
\newcommand{\earr}{\end{array}}
\newcommand{\beas}{\begin{eqnarray*}}
\newcommand{\eeas}{\end{eqnarray*}}

\title{
{\bf Nonlinear Dynamics of the Classical Isotropic Heisenberg Antiferromagnetic Chain : The
 Sigma Model Sector and the Kink Sector }}

\author{Radha Balakrishnan$^1$  and Rossen Dandoloff  $^2$}

\address{$^1$The Institute of Mathematical Sciences, Chennai  600 113,
India \\ 
 $^2$Laboratoire de Physique Th\'{e}orique et Mod\'{e}lisation,
 Universit\'{e} de Cergy-Pontoise, F-95031 Cergy-Pontoise, France.\\}

\abstract
{We identify {\it two distinct} low-energy sectors in the classical isotropic antiferromagnetic Heisenberg
 spin-S chain. In the continuum limit, we show that {\it two} types of rotation generators
 arise for the field in each sector. Using these, the  Lagrangian for sector I is  shown to
  be that of the nonlinear sigma model. Sector II  has a null Lagrangian;
  Its Hamiltonian density is just the Pontryagin term.  Exact solutions  are found in the form of 
  magnons and precessing pulses in I and moving kinks in II. The kink has `spin' S.
  Sector I has a higher minimum energy than II.} 
\maketitle2
\narrowtext
There are certain interacting many-body systems, which display a close correspondence between their
  extreme quantum behavior and strictly classical behavior, in the following sense: 
 If the {\it quantum} energy spectrum of the one-dimensional {\it lattice} Hamiltonian is  exactly solvable, then
 the {\it classical continuum} limit of its dynamical equation, which is usually 
a nonlinear partial differential equation,
  is completely integrable, supporting soliton solutions. As examples \cite{jevi} we note the correspondence between : 
  the quantum spin-$\frac{1}{2}$ isotropic ferromagnetic chain  and the classical
   Landau-Lifshitz equation; bosons with delta-function interactions and the nonlinear 
 Schr\"{o}dinger equation.

  Like its ferromagnetic (FM) counterpart in the example cited above, the  isotropic
 antiferromagnetic (AFM) Heisenberg chain described by the Hamiltonian 
\beq
 H \ = \ - J \sum_n {\bf S}_n \cdot {\bf  S}_{n+1},
\eeq
where ${\bf S}_n$ denotes the spin at the $n^{th}$ site,
 $({\bf S}_n)^2=S^2$ and $J < 0$,  is also  exactly solvable \cite{bethe} in the extreme quantum case $S=\frac{1}{2}$.
Faddeev and Takhtajan \cite{fadd} used Bethe ansatz techniques to show that the low-lying excitation
  here is a quantum kink with spin $\frac{1}{2}$. 
  Subsequently, using a formulation proposed by Mikeska \cite{mike1},
 Haldane \cite{hald1} showed that the classical continuum dynamics  of this
 system can be mapped to that of the $O(3)$ nonlinear sigma model (NLSM) Lagrangian, 
  whose  equations of motion  are known to be integrable \cite{pohl}. However,
   as he rightly pointed out, a {\it finite anisotropy} term must be added to the AFM Hamiltonian in (1),
    to obtain  a {\it classical}  kink solution (i.e., a topological soliton) for the spin configuration.
  (As is obvious from the functional form of this solution,  it ceases to be a kink in the isotropic limit.)
 Soon afterwards, the {\it quantum},  large spin $S$,  AFM chain was mapped \cite{hald2},\cite{affl} in the continuum to
  the Lagrangian of the NLSM {\it with} a topological term. Since this term is a `total divergence',
  the equations of motion are unaffected by it; and therefore, here too, the presence of an anisotropy
  appears to be crucial to support a kink.  Given the quantum$\leftrightarrow$classical
 correspondence mentioned in the beginning, and the rigorously-proven  existence of the quantum kink
 in the isotropic case, this apparent absence of the classical kink in the isotropic
 limit is strange indeed \cite{mike2}. This motivates us to take a deeper look into the classical equations
 of motion generated by the basic AFM Hamiltonian (1).

  In this Letter, in addition to showing how a kink  arises in the isotropic AFM  as an exact
 dynamical solution in the continuum,  we present several other
   new results that provide important  insights into the  low-energy dynamical structure underlying this system:
  On  the discrete lattice, we  identify  {\it two distinct}  low-energy sectors
  I and II that one must analyze for a full understanding of this system.  
  We derive the continuum 
  dynamical equations of (1), taking care to ensure that the Poisson bracket structure is preserved
  exactly. These equations help us to directly identify the corresponding total
 angular momentum in each sector. A careful scrutiny of the discrete system shows 
  that in the continuum limit, there are in fact {\it two}  types of rotation generators that get
  associated with the vector field (in each sector). When appropriately taken into account, these 
  yield the above total angular momentum. Using this, its Lagrangian
   is shown to be that of the NLSM. In this sector, we  obtain exact nonlinear
 magnons and precessing pulse solutions for the unit vector field.
 Sector II is described by a {\it single} equation for the vector field. 
 The associated Hamiltonian
  turns out to be a  homogeneous function of the  first degree in  the canonical momenta. Using 
 a Lagrangian analog of Dirac's \cite{dira} formulation of {\it constrained dynamics}, we show 
 that the Lagrangian of this sector vanishes identically.
  We find exact moving kinks for the field equation. This kink {\it does not}  require an anisotropy
 to support it, in contrast to the NLSM sector I. We show that the total  angular momentum
  of this kink is simply $S$. It carries a `vorticity' with it, reminiscent of the superfluid vortex.    
 Typically, the minimum energy of sector I is greater than that of sector II.
    
 We begin our analysis  by defining  
\beq
{\bf {n}}(2i+{1}/{2})= [{\bf S} (2i+1)-{\bf S}(2i)]/2S;
\eeq
\beq
{\bf {l}}(2i+{1}/{2})= [{\bf S} (2i+1) +{\bf S}(2i)]/2a. 
\eeq   
 Here $a$ is the lattice separation. Eqs. (2) and (3) are just the {\it classical}
 counterparts of the `staggered magnetization'  ${\bf n}$,
 and the `spin density' ${\bf l}$,  customarily used in the quantum formulation \cite{affl}. 
 The  dynamical equations for ${\bf S}(2i)$ and ${\bf S}(2i+1)$ are found from
 ${d{\bf S}_n}/dt \ = \ \{{\bf S}_n, H \}$, where $\{,\}$ denotes the Poisson bracket (PB).
 $S^\alpha_n$ satisfy the usual angular momentum algebra
$ \{S^\alpha_n , S^\beta_m\} \ = \ \delta_{nm} \epsilon^{\alpha \beta \gamma} S^\gamma_n$.
 Using this,  Eq.(1) yields ${d{\bf S}_n/dt} \ = \ J {\bf S}_{n} \times ({\bf S}_{n+1} +
{\bf S}_{n-1})$.
 In an AFM chain, the low-energy dynamics corresponds to  
$[{\bf S} (2i+1) +{\bf S}(2i)]$  being very small. In the continuum limit $a\rightarrow 0$,
  this leads to two possibilities : (I)   
$[{\bf S} (2i+1) +{\bf S}(2i)] = o(a^{m}), m\ge 1$, and  (II)   
$[{\bf S} (2i+1) +{\bf S}(2i)] = 0$, identically.   
Noting that the usual AFM spin wave velocity $c = 2|J|Sa$ must remain finite as $a\rightarrow0$,
 our detailed analysis \cite {BD} shows that in (I), only $m=1$ leads to nontrivial dynamics. 
 All $m\ge 2$ lead to   trivial spin configurations, constant in space and time. (II) corresponds to a `dimer-like'
 configuration in which every {\it second} pair of spins on the chain gets `locked' in 
 {\it  exactly} antiparallel directions, and has interesting dynamics. We proceed to analyze these two cases.

{\bf Sector (I)}: Here, ${\bf l}(2i+ \frac{1}{2})$  is nonzero. The two constraints 
 ${\bf n}^2=1+ (a/S)^{2} {\bf l}^{2}$ and ${\bf n}\cdot{\bf l}=0$ ensure that
 there are four independent variables per unit cell ($2a$), describing this sector  as required \cite{simon}.
 In the continuum limit $a \rightarrow 0$, we have, ${\bf n}^2 \rightarrow 1, (2i+\frac{1}{2})a \rightarrow x$,
 $\delta_{mn}/2a \rightarrow \delta(x-y)$, etc. Using the angular momentum algebra of the 
spins ${\bf S}_n$, one obtains for the continuum, 
$$
\{l^\alpha (x) , l^\beta (y)\} =\epsilon^{\alpha \beta \gamma} l^{\gamma}(x)\delta (x-y)\,\,\,;\,\,\, \{n^\alpha (x) , l^\beta (y)\}= $$
$$\,\,\,\,\,\,\,\,\,=\,\,\,\,\epsilon^{\alpha \beta \gamma} n^{\gamma}(x)\delta {(x-y)}\,\,\, \,;\,\,\,\, \{n^\alpha (x) , n^\beta (y)\} \rightarrow 0.
\,\,\,\,\,\,\,\,\eqno(4)   
$$ 

 Thus the spin density ${\bf l}$ behaves like an angular momentum for the unit vector field ${\bf n}$
 in the continuum limit. From the discrete equations of motion for  
 ${\bf S} (2i+1)$  and ${\bf S}(2i)$ arising from (1), we write down the equations for 
${\bf l}(2i+\frac{1}{2})$  and ${\bf n} (2i+\frac{1}{2})$,
 and   find their continuum limit by using appropriate Taylor expansions  for ${\bf l}(x\pm 2a)$
 and ${\bf n} (x\pm 2a)$. We obtain the following
 coupled equations :
$$
{\bf n}_t=(2c/S)[({\bf l}-\frac{S}{2}{\bf n}_x) \times {\bf n}] \eqno(5)
$$ 
$$
{\bf l}_t=c[({\bf l}-S{\bf n}_x) \times {\bf n}]_{x}. \eqno(6)
$$
The subscripts $x$ and $t$ stand for partial derivatives. Eq. (5) immediately yields 
$$
{\bf l} =(S/2)[\frac{1}{c}({\bf n}\times {\bf n}_{t}) +  {\bf n}_x] \eqno(7)
$$
On substituting Eq. (7) in Eq. (6), we get 
$$
{\bf n}\times [({\bf n}_{tt}/c^{2})-{\bf n}_{xx}] =0. \eqno(8)
$$
Eq. (8) is just the Euler-Lagrange equation for the NLSM, and is integrable \cite{pohl}.
 Using Eqs. (2) and (3) in (1), and taking the continuum limit,
$$
H = \ (c/S)\int dx [({\bf l}-\frac{S}{2}{\bf n}_x)^{2} +  (S^{2}/4) {\bf n}_{x}^{2})]=\int h  \ dx. \eqno(9)
$$
 To find the  Lagrangian density $\mathcal{L}$ for (9), we need to identify  the  {\it total}  canonical momentum
 ${\bf {{\pi}}}$. Now, an inspection of Eq. (5), which essentially  is an  equation of motion for a
   `nonlinear rigid rotator' ${\bf n}$, hints strongly that its  {\it total} angular momentum is in fact    
${\bf \Lambda}_{1}=({\bf l}-\frac{S}{2}{\bf n}_x)$ rather than just ${\bf l}$, which satisfies (4).
 To understand the  origin of this fact, we first define
$$
{\bf {L}}(2i+\frac{1}{2})=[\,{\bf S} (2i-1) +{\bf S}(2i)+{\bf S} (2i+1) +{\bf S}(2i+2)\,]/4a.\eqno(10) 
$$ 
This is an obvious extension of ${\bf l}$ defined in (3) and represents the spin density at $(2i+\frac{1}{2})$
 due to a cluster of $4$ spins around this point, with the appropriate denominator $4a$.
 A short calculation using the usual PB algebra obeyed by spins ${\bf S_n}$ shows that
 in the continuum limit, ${\bf L}$ also behaves like an angular
  momentum of ${\bf n}$, i.e., its components  satisfy (4). 
 Next, writing the expressions for the various spins appearing in (10) in terms of ${\bf n}$ and ${\bf l}$
 by using (2) and (3), and taking the continuum limit, we get
$$
{\bf L} ={\bf l}-S{\bf n}_x  \eqno(11)
$$
 Further, let us denote ${\bf l}$ in (3))  by ${\bf L}_2$, and ${\bf L}$ in (10) by ${\bf L}_4$, with the
 subscripts denoting the number of spins in the cluster, and similarly define spin densities     
${\bf L}_6, {\bf L}_8, {\bf L}_{10} $ etc., corresponding to clusters of $6,8,10$..spins. Then,  computing the 
 PBs of these ${\bf L}_n$ with ${\bf n}(2i+\frac{1}{2})$ shows that in the continuum limit,
  each of them satisfies Eq. (4), and so can qualify as the angular momentum of ${\bf n}$. In addition, a tedious 
 calculation shows that in the continuum, the spin densities corresponding to  ${\bf L}_6$, ${\bf L}_{10}$,
${\bf L}_{14}$ etc, reduce simply  to ${\bf l}$, while ${\bf L}_4$, ${\bf L}_{8}$, ${\bf L}_{12}$ etc,  
 reduce to ${\bf L}$ given in (11). Thus the contributions from the
 successive ${\bf L}_n$'s alternate between ${\bf l}$ and ${\bf L}$,
  leading to an {\it average} spin density $({\bf l+L})/2 $.  On using
  Eq. (11), this  indeed  gets identified with the the total angular momentum ${\bf \Lambda}_1$: 
$$
{\bf \Lambda}_1= ({\bf l}+{\bf L})/2  =  ({\bf l}-\frac{S}{2}{\bf n}_x) = (S/2c)({\bf n} \times {\bf n}_{t}),\eqno(12) 
$$ 
 where we have used Eq. (7) to write the last equality. 

To find the Lagrangian, ${\bf n}$ is written in spherical polar coordinates as
$$
{\bf n}= (\sin \theta \cos \phi, \sin \theta \sin \phi, \cos \theta)\eqno(13)
$$
Let $\pi_{\theta}$ and $\pi_{\phi}$ denote the canonical momenta satisfying 
 $\{\theta(x), \pi_{\theta}(y) \} =\{\phi(x),\pi_{\phi}(y)\} = \delta (x-y)$.  
 Noting that $({\bf n},{\bf n}_{\theta},{\bf n}_{\phi}(\sin \theta)^{-1}))$ 
form a unit orthogonal
 triad, we have  as usual \cite{affl}, ${\bf \Lambda}_1= ({\bf n}\times [\pi_{\theta}
{\bf n}_{\theta} +
 \pi_{\phi}{\bf n}_{\phi}(\sin \theta)^{-2})]$. Here, the subscripts $\theta$ and $\phi$ of ${\bf n}$
 denote partial derivatives. Comparing this with  the last equation in (12), we get    
 $\pi_{\theta} = (S/2c)\theta_{t}$ and $\pi_{\phi} = (S/2c) \sin ^{2}\theta  \phi_{t}$. Further,
  from  Eq. (9), $h=[{\pi_{\theta}}^{2}+{\pi_{\phi}}^{2}(\sin \theta)^{-2}+
  \sin^{2}\theta \phi_{x}^{2} + \theta_{x}^{2}] $.
 Using  these in $\mathcal{ L}$ =$ \pi_{\theta}\theta_t+\pi_{\phi}\phi_t -h$,
 we obtain $\mathcal {L}$=$[ (\frac{1}{c^{2}}){\bf n}_{t}^{2} - {\bf n}_{x}^{2} ].$
 This is just the NLSM Lagrangian. Note that
  {\it both} the rotation generators ${\bf l}$ and ${\bf L}$  must be used for the correct
  identification of the total  angular momentum of ${\bf n}$ in the continuum, which in turn is 
   needed to find the associated Lagrangian. This is a subtle point, indeed.
  Our detailed analysis shows that ignoring either one of these and treating the other alone as the
  total angular momentum  would  lead to a  {\it spurious} topological term  in the
  classical NLSM Lagrangian. 

  Analyzing Eq. (8) in terms of $\theta$ and $\phi$ variables using Eq. (13),
   we  find {\it exact} traveling wave solutions of the form $\theta = \theta(\omega_{1} t-k_{1}x)$
   and $\phi=(\omega t-kx)$ . Details \cite {BD} will be published elsewhere. For these
  solutions, the product of the velocities  of the $\theta$-wave and the $\phi$-wave  turns out to be
  $v_{\theta} v_{\phi}$ =$(\omega/k)(\omega_{1}/k_{1})$=$c^{2}$, implying that these waves travel
 in the same direction. As an example, we have the following `precessing pulse' that travels with velocity $c$:
$$
{\bf n}= (-tanh \, \xi\,\, \cos \xi, -tanh \, \xi\,\,  \sin \xi,\,\,  sech \,  \xi ),\eqno(14) 
$$
where $\xi=(\omega t - kx)$ and $\omega=ck$. As $|x| \rightarrow \infty $, although 
 $ n^{(3)} \rightarrow 0$, indicating a `pulse', $n^{(1)}$ and $n^{(2)}$ do not vanish, but precess on a plane. 
 We also find exact nonlinear analogs of the usual small amplitude spin wave solutions  of Eq. (8).
Typically, the traveling waves in this sector have a constant energy density.
 The consequences of this will be discussed at the end. 

{\bf Sector (II)}:
As already mentioned, here we have ${\bf S}(2i)=-{\bf S}(2i+1)$. Thus Eq.(2)
yields:
$$
{\bf n}(2i+1/2)\rightarrow {\bf S}(2i+1)/S=-{\bf S}(2i)/S
\equiv {\bf N}(2i+1/2)   \eqno(15)
$$
and Eq.(3) gives ${\bf l}(2i+\frac{1}{2})\equiv 0 $.
 From Eq.(15), it is easily verified that in the continuum limit
 $a\rightarrow 0$, $ \{ N^{\alpha}(x), N^{\beta}(x) \}\rightarrow \frac{2a}{S}\delta(x-y)\epsilon^
{\alpha\beta\gamma}N^{\gamma}(x)\rightarrow 0 $.
Thus {\bf N}(x) behaves like a unit vector field. 
 Using the dynamical equations for {\bf S}(2i) and {\bf S}(2i+1)
leads in the continuum limit to the following {\it single}  equation for {\bf N}:
$$
{\bf N}_t=c\,\,({\bf N}\times{\bf N}_x)  \eqno(16)
$$
Since ${\bf N}^2=1$, writing it in terms of polars $(\theta,\phi)$  as  we did for ${\bf n}$ in Eq. (13),
 leads to:
$$
\theta_t=-c\sin\theta\,\phi_x  \,\,\, ; \,\,\, \phi_t=c\,\theta_x/\sin\theta
  \eqno(17)
$$
Using Eq.(15) in Eq.(1) yields the following continuum Hamiltonian:
$$
H_0=\frac{cS}{2}\int dx ({\bf N}_x)^2 =\frac{cS}{2}\int dx ({\theta_x}^2+{\sin}^2\theta\,{\phi_x}^2)=$$
$$ = \int h_0\, dx  \eqno(18)
$$
 Proceeding as in sector I leads to  the total angular momentum
  ${\bf \Lambda}_{2}= -(S/2){\bf N}_{x}$ for this sector. Further,   
  ${\bf \Lambda}_2= ({\bf N}\times [p_{\theta}  {\bf N}_{\theta} +
 p_{\phi} {\bf N}_{\phi} (\sin \theta)^{-2})]$.  Writing ${\bf N}_{x}={\bf N}_{\theta} \theta_{x} 
 +{\bf N}_{\phi} \phi_{x}$, we compare the above two expressions of ${\bf \Lambda_{2}}$ 
 to obtain the momenta
$$
p_{\theta}=-(S/2) \sin  \theta\,  \phi_{x} \,\, ;\,\, p_{\phi}=(S/2) \sin \theta\, \theta_{x}.\eqno(19)
$$
Note the dependence of the momenta on the `generalized coordinates', $\theta$ and $\phi$  and their spatial derivatives.
 It is clear that to generate  the equations of motion (17), from
 $\theta_{t}= {\delta H_{0}}/\delta {p_{\theta}}$ and
  $\phi_{t}= {\delta H_{0}}/\delta {p_{\phi}}$, 
  the corresponding Hamiltonian density $h_{0}$  must have  the form
$$
h_{0}=-c\, [p_{\theta}\, \sin\theta \, \phi_{x} - (p_{\phi}/ \sin \theta)\, \theta_{x}] \eqno(20)
$$
 Thus  $h_{0}$ is  homogeneous of the  first degree in the $p$'s. 
 Note that there are no {\it independent}  equations for 
$(p_{\theta})_{t}$ and $(p_{\phi})_{t}$ in this sector.
A short calculation shows that the corresponding Lagrangian density $\cal {L}_O$=$  
p_{\theta}\theta_{t} + p_{\phi} \phi_{t} -h_{0}$ {\it vanishes} identically. This may be surprising at first
  sight, but not if it is realized that the dynamics in this sector is just an example
 of the Lagrangian analog of Dirac's \cite{dira} constrained dynamics, encountered in particle mechanics.
 Treating  the dynamical equations (16) as constraints, which leads to ${\bf N}_{t}^{2}/c^{2} ={\bf N}_{x}^{2}$,
 our null Lagrangian may be written in the form $\mathcal L_{O}$=$({\bf N}_{t}^{2}/c^{2} -{\bf N}_{x}^{2})$,
  which yields $p_{\theta}=(S/2c) \theta_t$ and $p_{\phi}=(S/2c) \sin^{2}\theta \phi_{t}$. These yield Eq.(19),
   on using Eq. (17). Further, using these in (20) gives 
 $h_{0}=(S/2) \sin \theta [\theta_{x}\phi_{t}-\theta_{t}\phi_{x}]= (S/2) {\bf N}\cdot
 ({\bf N}_{x}\times {\bf N}_{t})$. Thus the energy density in this sector is effectively  the Pontryagin
 (topological) term. To our knowledge, this is the first
example in condensed matter physics where Dirac's method for constrained 
dynamics finds an explicit application.  

   We   have found \cite{BD} {\it exact} traveling wave solutions of the form $\theta = \theta(\omega_{1} t-k_{1}x)$
   and $\phi=(\omega t-kx)$  for the dynamical equations (17) in this sector. In general these
  are traveling kinks, for which  the product $v_{\theta} v_{\phi}$ =$(\omega/k)(\omega_{1}/k_{1})$=$-c^{2}$,
  implying that these two waves travel in {\it opposite} directions to each other, in contrast to sector I.
  As an example, we present the following solution:
$$
{\bf N}= (sech\, \xi_{1} \,\cos \xi, sech \, \xi_{1} \, \sin \xi, -tanh \,  \xi_{1} ).\eqno(21) 
$$
Here $\xi_{1}=(\omega t + kx )$,  $\xi = (\omega t - kx)$  and $\omega = ck$. 
 Since ${\bf N}={\bf S}(2i+1)/S={\bf -S}(2i)/S$, Eq. (21)  represents a kink profile in each sublattice,
 moving with velocity $c$, with the spins precessing around the z-axis with velocity
 $-c$. The odd site spins rotate from $\theta = 0 $ at $x \rightarrow -\infty$ to $\theta = \pi$
 at $x \rightarrow +\infty$, and vice versa for the even  site spins.  Unlike in sector I,
  the  kink is a  topological soliton  that interpolates between
 two {\it distinct} N\'{e}el states.
  Typically, the traveling wave solutions in this sector have an
 energy density which is a  `lump', vanishing as $|x| \rightarrow \infty$. The total energy of the
 kink is found from Eq. (18) to be
 $E_{K}= 2 S \omega = 2 S c k = 2Sc/\Gamma$, where $\Gamma=(1/k)$ is the width of the kink. Thus the energy vanishes
  as $k \rightarrow 0$. The z-component of the integrated total angular momentum of the kink  is $M_{K}^{(3)}=
  \int \Lambda_{2}^{(3)}  dx = (-S/2)\int N_{x}^{(3)} dx = S$, while  $M_{K}^{(i)}$, $i=1,2$ vanish. 
Thus the total angular momentum of the kink is `self-generated' and can hence be interpreted as its `spin'. Its
magnitude  is simply the spin S of the chain (1).  Further, there is a clear spatial separation between the
  `vorticity' field ${\Lambda_2}^{(3)}(x,t)=\frac{Sk}{2}{sech}^2(kx+\omega t)$ which is a pulse,  and the
  kink field $N^{(3)}(x,t)=-tanh(kx+\omega t)$, as in a superfluid. 

The above results correspond to $[{\bf S}(2i)+{\bf S}(2i+1)]=0$, implying
${\bf l}(2i+\frac{1}{2})=0$ (see Eq.(3)). If we consider the other possibility
$[{\bf S}(2i)+{\bf S}(2i-1)]=0$, then clearly ${\bf L}(2i+\frac{1}{2})=0$ (see Eq.(10)).
Repeating the same steps as before, we obtain ${\bf N}_t =-c  ({\bf N}\times {\bf N}_x)$,
 This implies waves traveling with velocity c {\it opposite} in direction to that
in Eq.(16 ).
  Further,  ${\bf l}=S{\bf N}_x$ here,
 yielding ${\bf \Lambda}_{2}= (S/2){\bf N}_x$. 
 Thus Sector II supports kink solutions with
velocity $\pm$ c depending on which of the `dimer' bonds is in the antiparallel-locked configuration.
These kinks have identical energies, total angular  momenta etc. 

Returning to Sector I, we have shown that the {\it isotropic} AFM chain maps
to the NLSM model {\it without} a topological term. This  appears to be
 in agreement with the result that Haldane \cite{hald1} had obtained in the
  isotropic limit, in his 
seminal paper using Mikeska's ansatz \cite{mike1}. It is therefore instructive to check if this ansatz, namely,
${\bf S}_n = (-1)^nS(\sin\theta_n\cos\phi_n, \sin\theta_n\sin\phi_n, 
\cos\theta_n)$, with $\theta_n=\theta(x)+a(-1)^n\alpha(x)$, $\phi_n=
\phi(x)+a(-1)^n\beta (x)$, $x=na$  satisfies the basic relationship
${\bf l}=\frac{s}{2}[\frac{1}{c}{\bf n}\times{\bf n}_t + {\bf n}_x]$ (see Eq.(7))
derived from the {\it exact} continuum equations  of (1)). A long calculation
shows that $\alpha$ and $\beta$ must equal $({\theta_x}/{2})$ and 
$({\phi_x}/{2})$ respectively, to satisfy the above relationship. This in turn
 can be  shown to imply that the ansatz is indeed consistent for {\bf l}=0 (Sector II) and
therefore leads to  our Eq. (17), which supports kink solitons.
 Using this equation, the Lagrangian (Eq.(10) in $\cite{hald1})$ is easily seen to
  {\it vanish} in the isotropic limit, as expected. 

 Eq. (17)  also supports  instanton solutions, since it is just the (1+1)D analog of the 2D 
Belavin-Polyakov equation \cite{bela}.
This equation has been shown to be integrable by mapping it to the elliptic 
Liouville equation \cite{rb}. Thus the NLSM equation (Eq.(8)) of sector I
 and  the Belavin-Polyakov equation (Eq.(16)) of sector II are both integrable.

 Finally, we discuss the energy associated with the two sectors: For sector I, a short calculation
 using Eq. (9) shows that the minimum energy density corresponds to ${\bf l}_{min}=(S/2) {\bf n}_x$
    yielding  $h_{min} = (c/S) {\bf l}_{min}^{2}$.  Recalling the restriction  
  that  both generators of rotation, ${\bf l}$ and ${\bf L}$ have to be finite in this sector, 
 leads to the result that  $h_{min}$ is {\it finite} in this sector.
 In contrast, for sector II, from Eq. (18), the minimum energy density is directly seen to be $h_{0,min}=0$,
 corresponding to ${\bf N}_x=0$, implying that it is continuously connected to the ground state energy.
  Thus sector I has a higher minimum energy than sector II.   

 All our results so far have been fully classical.  Although a rigorous quantum mechanical analysis
 of the two sectors is needed for the full understanding of the energy spectrum, 
 {\it semiclassically}, we see that for sector I, our classical result  
   implies $E_{min}\sim l(l+1)$. Since  we have the restriction $l\neq0$, only the next higher
 integer value  of $l$ is allowed, leading to a gap. 
 On the other hand, there is no such restriction on either of the  angular momenta  in sector II, and
  hence   $E_{min}=0$ here.  
 Thus at any finite temperature, the ground state of the isotropic  AFM chain gets
 disordered due to the excitation of kinks  rather than  magnons \cite{hald1}. 
 Further, as we showed earlier, the spin of the kink is $S$. Specializing to the $S=\frac {1}{2}$ case,
 our analysis suggests that a  kink  in the isotropic AFM chain  is a  spin-$\frac{1}{2}$ entity 
 which is gapless, in agreement with Faddeev and Takhtajan's result \cite{fadd} as well.

R.B  and R.D would like to thank, respectively, the Universit\'{e} de Cergy-Pontoise and {\it CNRS}, France,
  and  the Institute of Mathematical Sciences, Chennai, India,  for visiting professorships which
 made this collaboration possible.


\begin{thebibliography}{99}
\bibitem{jevi} A. Jevicki and N. Papanicolau, Ann. Phys. (N.Y.) {\bf 120}, 107 (1979); 
 C. R. Nohl, Ann. Phys. (N. Y.) {\bf 96}, 234 (1976). 
\bibitem{bethe} H. A. Bethe, Z. Phys. {\bf 71}, 205 (1931). 
\bibitem{fadd}  L. D. Faddeev and L. A. Takhtajan, Phys. Lett. A {\bf 85}, 375 (1981).
\bibitem{mike1} H. -J. Mikeska, J. Phys. C  {\bf 13},  2913 (1980) .
\bibitem{hald1} F. D. M. Haldane, Phys. Rev. Lett. {\bf 50}, 1153 (1983).
\bibitem{pohl} K. Pohlmeyer, Commun. Math. Phys. {\bf 46}, 207 (1976).
\bibitem{hald2} F. D. M. Haldane, J. Appl. Phys. {\bf 57}, 3359 (1985).  
\bibitem{affl}  I. Affleck, Nucl. Phys. B {\bf 257}, 397 (1985); See also, J. Phys : Condens. Matter
 {\bf 1}, 3047 (1989).
\bibitem{mike2} For an excellent review of solitons in magnetic chains, see H. -J. Mikeska and M. Steiner, Adv. Phys.
 {\bf 40}, 191 (1991), especially pages 335 and 345.
\bibitem{dira}  P. A. M. Dirac,  {\it Lectures in Quantum Mechanics} (Belfer Graduate School of Science,
  Yeshiva University, New York 1964).
\bibitem{BD}  Radha Balakrishnan and R. Dandoloff (unpublished).
\bibitem{simon} S. Villain-Guillot and R. Dandoloff, J. Phys. A: Math. Gen. {\bf 31}, 5401 (1998). 
\bibitem{bela} A. Belavin and A. M. Polyakov, JETP Lett. {\bf 22}, 245 (1975). 
\bibitem{rb} Radha Balakrishnan, Phys. Lett. A {\bf 204}, 243 (1995).
\bibitem{fn} To create even small amplitude spin waves  in an AFM chain,
 the neighboring  spins must precess with unequal cone angles, implying  a finite ${\bf l}\neq 0$.
 See, e.g.,F. Keffer, H. Kaplan and Y. Yafet, Am. J. Phys. {\bf 21}, 250 (1953).

\end{thebibliography}
\end{document}